\documentclass[manuscript]{aastex}
\usepackage{booktabs}
\usepackage{natbib}
\shorttitle{Neutrino Transients in IceCube-DeepCore}
\shortauthors{IceCube et al.}

\begin{document}

\title{Search for Transient Astrophysical Neutrino Emission with IceCube-DeepCore}

\author{
IceCube Collaboration:
M.~G.~Aartsen\altaffilmark{1},
K.~Abraham\altaffilmark{2},
M.~Ackermann\altaffilmark{3},
J.~Adams\altaffilmark{4},
J.~A.~Aguilar\altaffilmark{5},
M.~Ahlers\altaffilmark{6},
M.~Ahrens\altaffilmark{7},
D.~Altmann\altaffilmark{8},
T.~Anderson\altaffilmark{9},
I.~Ansseau\altaffilmark{5}
% ...
}
\altaffiltext{1}{Department of Physics, University of Adelaide, Adelaide, 5005, Australia}
\altaffiltext{2}{Technische Universit\"at M\"unchen, D-85748 Garching, Germany}
\altaffiltext{3}{DESY, D-15735 Zeuthen, Germany}
\altaffiltext{4}{Dept.~of Physics and Astronomy, University of Canterbury, Private Bag 4800, Christchurch, New Zealand}
\altaffiltext{5}{Universit\'e Libre de Bruxelles, Science Faculty CP230, B-1050 Brussels, Belgium}
\altaffiltext{6}{Dept.~of Physics and Wisconsin IceCube Particle Astrophysics Center, University of Wisconsin, Madison, WI 53706, USA}
\altaffiltext{7}{Oskar Klein Centre and Dept.~of Physics, Stockholm University, SE-10691 Stockholm, Sweden}
\altaffiltext{8}{Erlangen Centre for Astroparticle Physics, Friedrich-Alexander-Universit\"at Erlangen-N\"urnberg, D-91058 Erlangen, Germany}
\altaffiltext{9}{Dept.~of Physics, Pennsylvania State University, University Park, PA 16802, USA}
% ...

\begin{abstract}
We present the results of a search for astrophysical sources of brief transient neutrino emission using IceCube and DeepCore data acquired between May 15th 2012 and April 30th 2013. While the search methods employed in this analysis are similar to those used in previous IceCube point source searches, the data set being examined consists of a sample of predominantly sub-TeV muon neutrinos from the Northern Sky (-5$^{\circ} < \delta <$ 90$^{\circ}$) obtained through a novel event selection method. This search represents a first attempt by IceCube to identify astrophysical neutrino sources in this relatively unexplored energy range. The reconstructed direction and time of arrival of neutrino events is used to search for any significant self-correlation in the dataset. The data revealed no significant source of transient neutrino emission. This result has been used to construct limits at timescales ranging from roughly 1\,s to 10 days for generic soft-spectra transients. We also present limits on a specific model of neutrino emission from soft jets in core-collapse supernovae.
\end{abstract}


\section{Introduction}
The nascent field of high-energy neutrino astronomy opens the possibility of answering several open questions in astrophysics due in large part to the neutrino's ability to escape the densest regions of astrophysical environments. Specifically, the detection of transient astrophysical neutrino sources will help shed light on the acceleration mechanisms at work in some of the most energetic phenomena in the Universe such as gamma-ray bursts, supernovae, and active galactic nuclei. Previous attempts to detect such sources with the IceCube Neutrino Observatory \citep{2006APh....26..155I} are most sensitive to neutrino fluxes above 1 TeV with poor sensitivity below 100 GeV. Searches for astrophysical sources at lower energies (1--100 GeV) have been performed by Super-Kamiokande \citep{2009ApJ...704..503T}, however the detector's 50 kton instrumented volume limits its sensitivity to astrophysical neutrino fluxes. A newly developed 30--300 GeV muon neutrino sample collected by IceCube and its low energy extension DeepCore \citep{2012APh....35..615A} enhances IceCube's sensitivity in this under-explored energy range. In this paper we will present the results of a search for transient neutrino emission in this GeV-scale neutrino sample.

The detection of astrophysical neutrino sources is a primary design goal of the IceCube Neutrino Observatory \citep{2006APh....26..155I}. Located at the geographic South Pole, IceCube utilizes the clear Antarctic glacial ice ice cap as a detection medium for the Cherenkov light produced by secondary products of neutrino interactions. The detector consists of 5,160 Digital Optical Modules (DOMs) distributed among 86 cables or ``strings" to form a 1 km$^3$ instrumented volume. These DOMs house photomultiplier tubes (PMTs), to detect Cherenkov photons, as well as digitizing electronics for initial processing of the PMT data \citep{2009NIMPA.601..294A}. A centrally located region of denser instrumentation featuring DOMs with more sensitive PMTs comprises the DeepCore sub-array. This extension to the IceCube array enhances the detector's response to lower energy neutrino events.

Typical searches for astrophysical sources with IceCube make use of a sample primarily comprised of an irreducible background of high-energy atmospheric muon neutrinos ($E_{\nu} \gtrsim 1$ TeV) to look for both steady \citep{2014ApJ...796..109A} and transient sources \citep{0004-637X-807-1-46}. As of yet, these searches have not found any significant self-correlations within the data sample nor correlations between the neutrino data and known astrophysical objects of interest. So far, these analyses have largely eschewed low energy neutrino events collected by DeepCore for two reasons. First, the poorer angular resolution of these events renders them less suitable for pointing analyses. Second, the soft spectrum of the atmospheric neutrino flux results in higher rate of background neutrino events. However, the increased background can be somewhat mitigated by searching solely for transient sources. Therefore, applying previously developed search techniques \citep{2010APh....33..175B} to a sample of low energy (30 GeV $\leq E_{\nu} < 300$ GeV) muon neutrino events from DeepCore can enhance IceCube's sensitivity to short transient neutrino sources with softer spectra.

Due to the large atmospheric neutrino background in this energy range, searches using a data set composed of these low energy events will only be sensitive to emission timescales on the order of one day or shorter. Active galactic nuclei (AGN) undergoing flaring events are one potential source for emission on this timescale. Protons may be accelerated in relativistic jets, powered by accretion onto the AGN, resulting in the production of pions (and subsequently neutrinos) in shocks due to proton-photon interactions and proton self-collisions \citep{2009APh....31..138B}. For some of the timescales under consideration in this search, AGN-powered hadron acceleration must occur over a compact region and will require very large acceleration gradients \citep{2013ApJ...779..106K}. The presence of these large gradients will result in significant acceleration of muons prior to decay, leading to spectral hardening of the neutrino flux. Thus, if neutrino emission is occurring over short timescales, it will feature enhanced visibility at higher energies.

Sub-photospheric neutrino emission from gamma-ray bursts (GRBs) represents another possible source for this search. A model for photospheric gamma-ray emission in GRBs by \cite{2013PhRvL.111m1102M} suggests that a substantial flux of 100 GeV-scale neutrinos may be produced during the initial stages of relativistic outflow in the GRB. Decoupling of protons and neutrons during the initial formation of the relativistic jet causes hadronuclear collisions resulting in the production of pions and the production of neutrinos via pion decay. The predicted energy for the neutrinos produced in these sub-photospheric collisions is on the order of 100 GeV, and therefore this GRB neutrino flux may only be visible to IceCube searches with the inclusion of sub-TeV neutrino events. 

Perhaps the most promising potential source for this study is a special class of core-collapse supernova referred to as choked GRBs \citep{2001PhRvL..87q1102M}. The standard GRB model assumes that relativistic jets are generated during the accretion of material onto the compact object formed during core-collapse \citep{1992MNRAS.258P..41R}. Fermi-acceleration of charged particles occurs within the internal shocks of these jets leading to gamma ray emission once the jets breach the surrounding stellar envelope. There is an observed correlation between long duration GRBs and core-collapse supernovae (CC SNe) (\citep{2006ARA&A..44..507W}, \citep{2011AN....332..434M}). While the observed fraction of SNe resulting in the occurrence of a GRB is quite low, it may be that a larger fraction of core-collapse SNe still manage to produce mildly relativistic jets.  Due to insufficient energy, these jets fail to break through the stellar envelope and any gamma ray emission is effectively `choked' off. If protons are accelerated in these jets, then neutrino production will occur in the shocks of the jet irrespective of whether or not the jet successfully escapes. A model of this neutrino emission proposed by \cite{2004PhRvL..93r1101R} and extended upon by \cite{2005PhRvL..95f1103A}, hereafter referred to as the RMW/AB model, suggests that these neutrinos may be detectable by IceCube-DeepCore for nearby supernovae \citep{PhysRevD.81.083011}. Previous IceCube analyses have investigated the RMW/AB emission model with respect to a specific source \citep{2011A&A...527A..28I} and as part of the optical followup program \citep{2012A&A...539A..60A}, however the presented search marks the first use of low-energy muon neutrino events in constraining this model.

We present the results of a search for transient neutrino emission with a set of low-energy neutrino event data collected from May 15th, 2012 to April 30th, 2013. The data selection methods used to acquire this unique event sample will be detailed in Sec. 2. Analysis methods and search techniques are discussed in Sec. 3. Finally, the results of the search are given in Sec. 4 in addition to how these results may be interpreted within the context of generic neutrino flares as well as choked GRBs under the RMW/AB model.
\section{Event Selection}

The data acquisition process begins with the fulfillment of one of three trigger conditions that prompt the readout of the detector data. Each of these triggers requires some number of DOMs to exhibit hard local coincidence (HLC) within a defined time window. To satisfy the HLC condition, two or more neighboring (or next-to-nearest-neighboring) DOMs on the same string must register photon hits within a $\pm1$ $\mu$s window. The trigger for the lowest energy events (often referred to as simple majority trigger 3 or SMT3) requires three HLC DOM hits within a time window of 2.5 $\mu$s among the DeepCore string DOMs (or in DOMs on IceCube strings neighboring DeepCore). The two other triggers that serve as input for this event selection operate over the entire detector array with one requiring eight HLC DOM hits in a 5 $\mu$s window (SMT8) and the other requiring four HLC DOM hits within a cylinder of height of 75m and a radius of 175m in a 1 $\mu$s window (Cylinder Trigger).

Events satisfying these trigger conditions are then passed to the DeepCore data filter \citep{2012APh....35..615A}. This filter reduces the number of cosmic ray muons by using the outer regions of the detector as an active veto to tag down-going events originating outside the detector. Specifically, the filter examines timing and position information of DOM hits inside the DeepCore fiducial volume to identify a center of gravity (CoG) or vertex. For each DOM hit in the veto region, the speed of a hypothetical particle connecting that veto region hit to the CoG inside the fiducial volume is calculated. Veto regions hits whose speed lies within a range consistent with that of the speed of light are causally related and are therefore likely the product of background cosmic ray muons. Events having more than one correlated veto region hit are removed by the filter.

During the observation period of this search, the DeepCore filter consisted of two separate branches characterized by differing definitions of fiducial and veto volumes as opposed to the single definition given in \cite{2012APh....35..615A}. Another key difference of the applied filter, with respect to the definition provided in \cite{2012APh....35..615A}, is that it now makes use of some isolated DOM hit information instead of only using HLC hits. Events satisfying the SMT3 trigger feed the standard DeepCore filter branch whose fiducial and veto region definitions are roughly equivalent to those described in \cite{2012APh....35..615A}. The SMT8 and Cylinder Trigger events, in addition to SMT3 events that fail the standard filter branch, feed into the other branch of the filter which makes use of a more relaxed veto region, consisting of two instead of three layers of IceCube strings, providing a larger detection volume. The output of both branches of this filter are used in this search with the standard three-layer veto focusing on low-energy events and the two-layer veto branch retaining higher energy events. These branches are referred to as the low-energy stream (LES) and high-energy stream (HES) and have an exclusive event rate of 17.3 Hz and 23.3 Hz, respectively.

\subsection{Veto Cuts and Event Reconstruction}
Events belonging to both the LES and HES are subjected to several cuts that make use of veto region hit information, event topology, and event reconstructions to reduce the volume of cosmic ray background events as well as eliminate events that are the result of PMT dark noise-induced triggering. The first of these cuts requires at least two DeepCore DOM hits within a 250 ns window to remove SMT3 events that are the result of spurious hits. An algorithm designed to search for track-like events is then used to eliminate noise-induced events that show little evidence of correlation in DOM hits. Additionally, events are required to have at least 10 hit DOMs, to allow for a well-constrained reconstruction. The DeepCore filter algorithm is also reapplied several times using looser DOM hit cleaning settings to allow more isolated DOM hits in the veto region to contribute to the vertex calculations. Finally, the number of DOM hits that occur prior to the first hit inside the DeepCore detection volume is used as a cut parameter to eliminate potential cosmic ray muon events missed by the filter.

The initial event reconstruction uses a simple linear fit \citep{2014NIMPA.736..143A} to determine the first-guess direction of a muon track that describes the observed DOM hit pattern. This linear reconstruction is then used as a seed for a likelihood-based reconstruction (\cite{2004NIMPA.524..169A}) which uses a single-photoelectron (SPE) hypothesis to describe the probability of DOMs receiving light from the track at a given time due to scattering in the ice. Six iterations of the SPE likelihood reconstruction are performed to obtain a best-fit track for the event. Any event with a reconstructed direction, from either the linear or SPE likelihood fit, more than 5$^{\circ}$ above the horizon is removed from the sample. We also require that the angular separation between these two reconstructions is less than 30$^{\circ}$ for events in the HES sample.

Spurious DOM hits that occur in the central detector prior to the arrival of cosmic ray muons allow many background events to elude detection through the standard veto technique. To isolate these events, a separate SPE likelihood reconstruction is performed without using any information from the first two DOM hits in the event. Just as before, events with a reconstructed direction more than 5$^{\circ}$ above the horizon are removed. Events in the LES portion of the sample are disproportionately affected by noise hits due to both the lower light yield of these events as well as the increased noise rate of the higher quantum efficiency DeepCore DOMs. An additional SPE likelihood reconstruction is performed for LES events that attempts to mitigate the noise contribution to the likelihood by requiring isolated DOM hits to be more strongly correlated to hits satisfying the HLC condition. Once again, if the best-fit direction from this additional reconstruction on LES events is 5$^{\circ}$ above the horizon, the event is removed.

A final event reconstruction uses the previously mentioned six iteration SPE likelihood fit as its seed. This reconstruction differs from the seed in two important ways. First, it uses a multi-photoelectron likelihood (MPE) instead of the simpler SPE algorithm used previously (see \cite{2004NIMPA.524..169A}). Second, a parameterization of Monte Carlo simulation of photon transport is used in place of an analytic approximation to model the timing distribution for the arrival of Cherenkov photons to the DOM PMTs \citep{2013CoPhC.184.2214W}. This reconstruction is identical to that used in a multi-year point source search with IceCube \citep{2014ApJ...796..109A} and the results of this fit are used for the final data analysis. In order to estimate the angular uncertainty of the reconstruction, the likelihood space about the reconstructed direction is fit with a paraboloid via the method described in \cite{2006APh....25..220N}. The angular uncertainty derived from the paraboloid method serves as event quality parameter, and only events having an estimated angular error $\sigma_i$ less than 45$^{\circ}$ are kept.
 
\subsection{Boosted Decision Tree}
After the application of the described veto and reconstruction cuts, the ability to separate the muon background from potential neutrino signal events via simple cuts is drastically reduced. We therefore use a boosted decision tree or BDT \citep{hastie01statisticallearning} in order to isolate a final sample with acceptable neutrino purity, i.e. $<10 \%$ of events are the result of background cosmic ray muons. The use of a BDT allows for the classification of events by examining several event parameters, and it is a technique that has proven useful in previous IceCube analyses \citep{2013ApJ...779..132A}. The decision tree is formed through an iterative process in which a series of cuts on event parameters is chosen to maximize separation between signal and background training samples. Any events that are misidentified by the decision tree are then re-weighted or `boosted' to increase the likelihood of correct classification by the next iteration of the decision tree.

At this level of event selection, the large majority of experimental data still consists of background cosmic ray muons allowing the actual data to serve as a background training sample for the BDT. Simulated muon neutrino events, generated by GENIE \citep{2010NIMPA.614...87A} and Neutrino Generator (a modified version of ANIS \citep{2005CoPhC.172..203G} specific to IceCube), are used for signal training of the BDT. Neutrino signal events belonging to the LES or HES branches exhibit significant differences in the distribution of the input BDT parameters, described below, necessitating the construction of two separate BDTs.

The event parameters used for the LES tree include the location of the reconstructed event vertex, the number of `direct' DOM hits (featuring a photon travel time residual between -25 and 150 ns with respect to the reconstructed muon track), the reduced log-likelihood of the MPE reconstruction, the average distance between DOM hits and the reconstructed track weighted by DOM PMT charge, and the highest clustering of veto region PMT charge (found by brute force reconstruction methods). The HES BDT makes use of the direct hits parameter described above, the reduced log-likelihood of the MPE reconstruction, the average charge-weighted DOM distance to track, and the best fit track length using information from direct DOM hits. A simulated signal neutrino event sample weighted to a E$^{-2.5}$ (LES) or E$^{-2}$ (HES) spectrum is used for signal training.
\begin{figure}[ht]
\plottwo{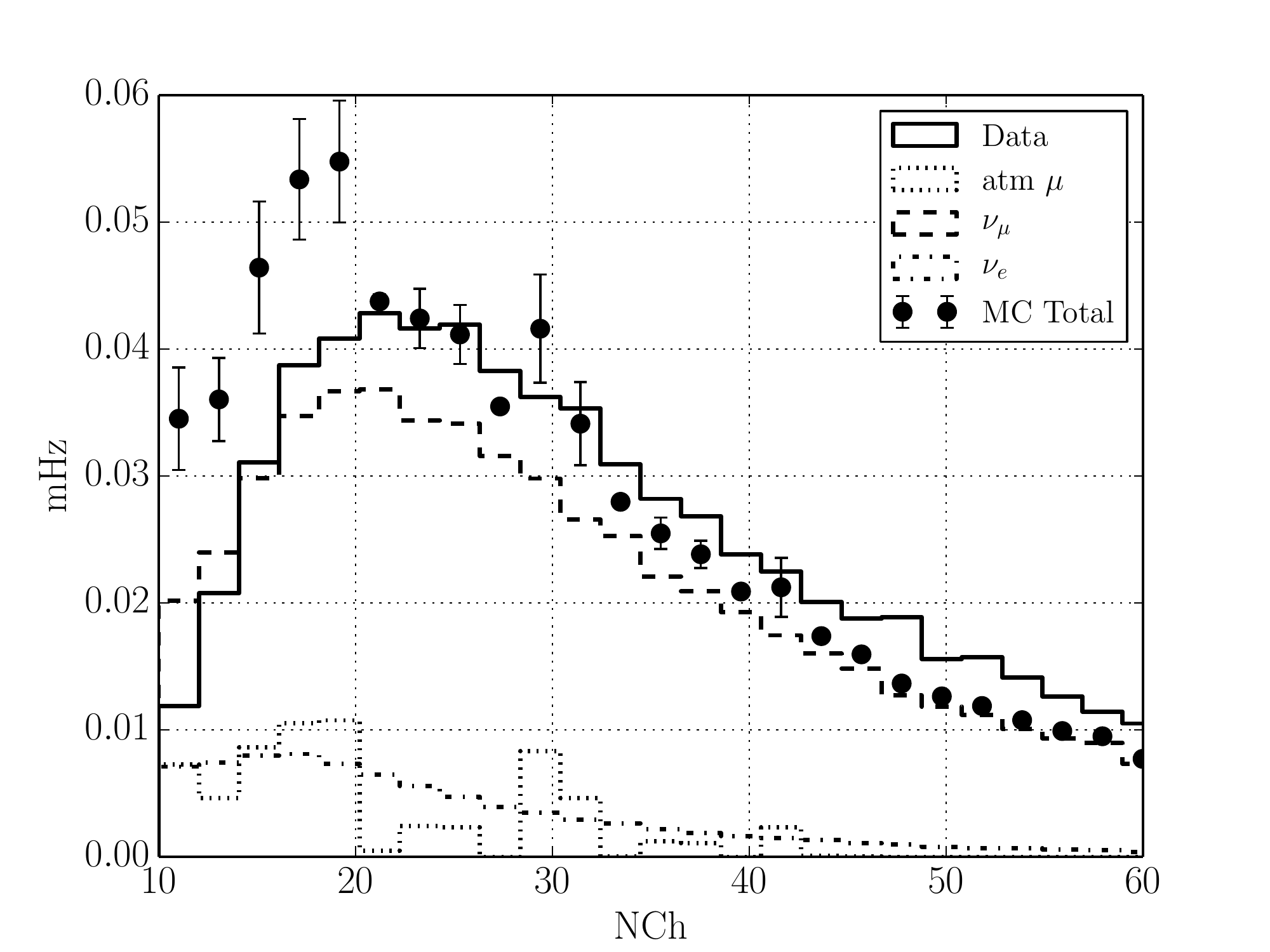}{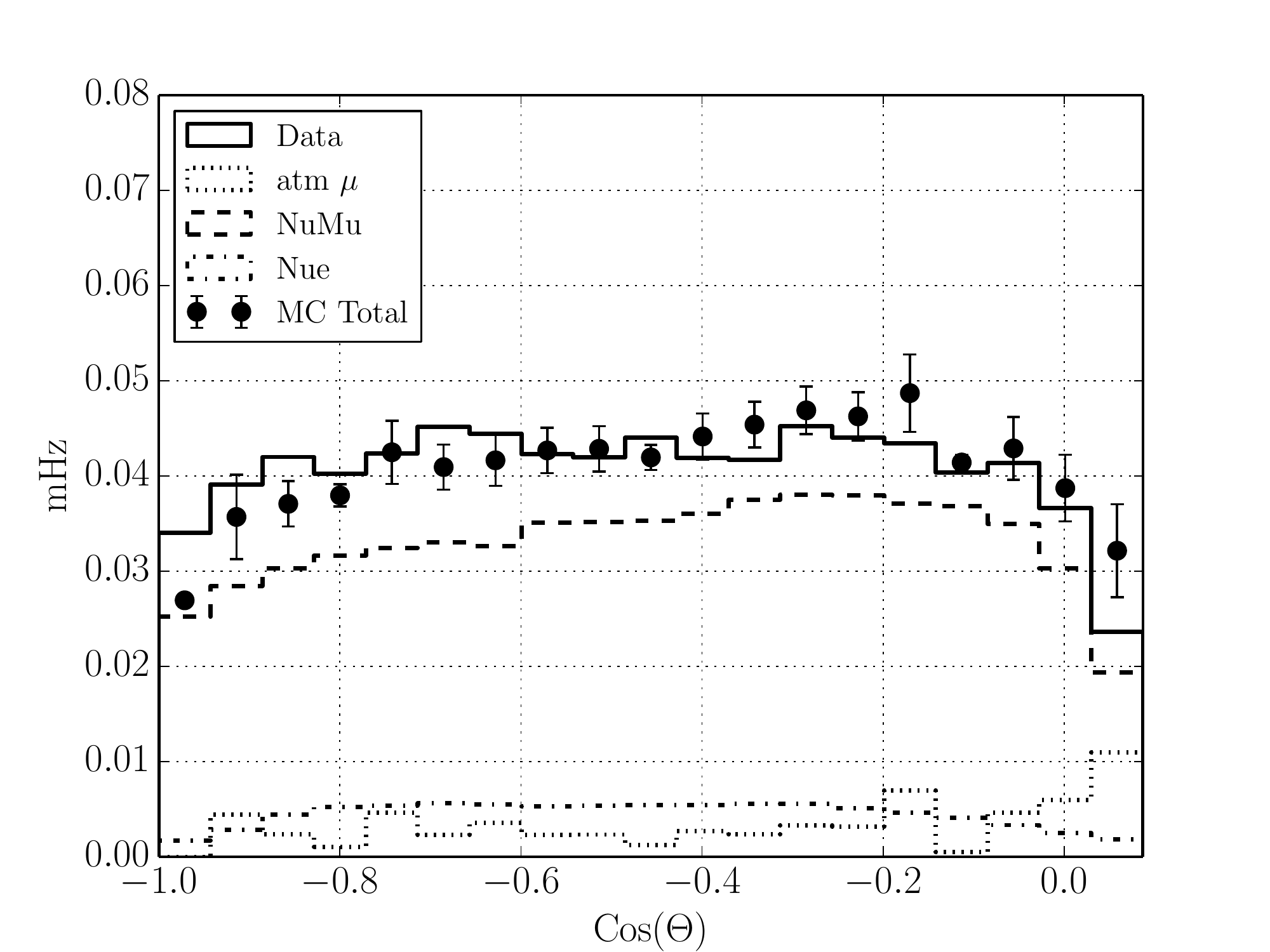}
\caption[Final Sample Event Parameter Distributions]{Final event rate distributions for the number of DOMs registering hits during the event (left) and the cosine of the reconstructed event zenith in detector coordinates (right). The solid line describes the final level dataset while the black points represent the sum of the various simulated species. Bins featuring large error are the result of atmospheric muon events, generated by CORSIKA \citep{1998cmcc.book.....H}, which suffer from limited statistics at the final level.}
\label{fig:PostBDTDistributions}
\end{figure}

Events are then input to the trained BDT, and a cut on the event BDT score is imposed to yield a data sample featuring a neutrino purity of approximately 90$\%$. This final event sample consists of 22,040 events over a livetime of $\sim$330 days, corresponding to a data rate of about 0.77 mHz. As Figure \ref{fig:PostBDTDistributions} indicates, the final sample is mostly composed of atmospheric neutrinos with an estimated cosmic ray muon contamination of approximately 0.07 mHz. There is a disagreement between simulation predictions and experimental data in the rate of events featuring a low number of DOM hits. The source of this discrepancy is not fully understood, however, the distributions of other event parameters, e.g. reconstructed zenith shown in Figure \ref{fig:PostBDTDistributions}, are well-described by the atmospheric neutrino simulation. Given the agreement between simulation and data for event parameters relevant to the analysis method, we contend that the simulation of signal events for the purpose of calculating the sensitivity of the search is accurate. Additionally, we do not rely on simulation for modeling of analysis background, and we instead use the experimental data itself to directly determine the background characteristics.

\begin{figure}[ht]
\plottwo{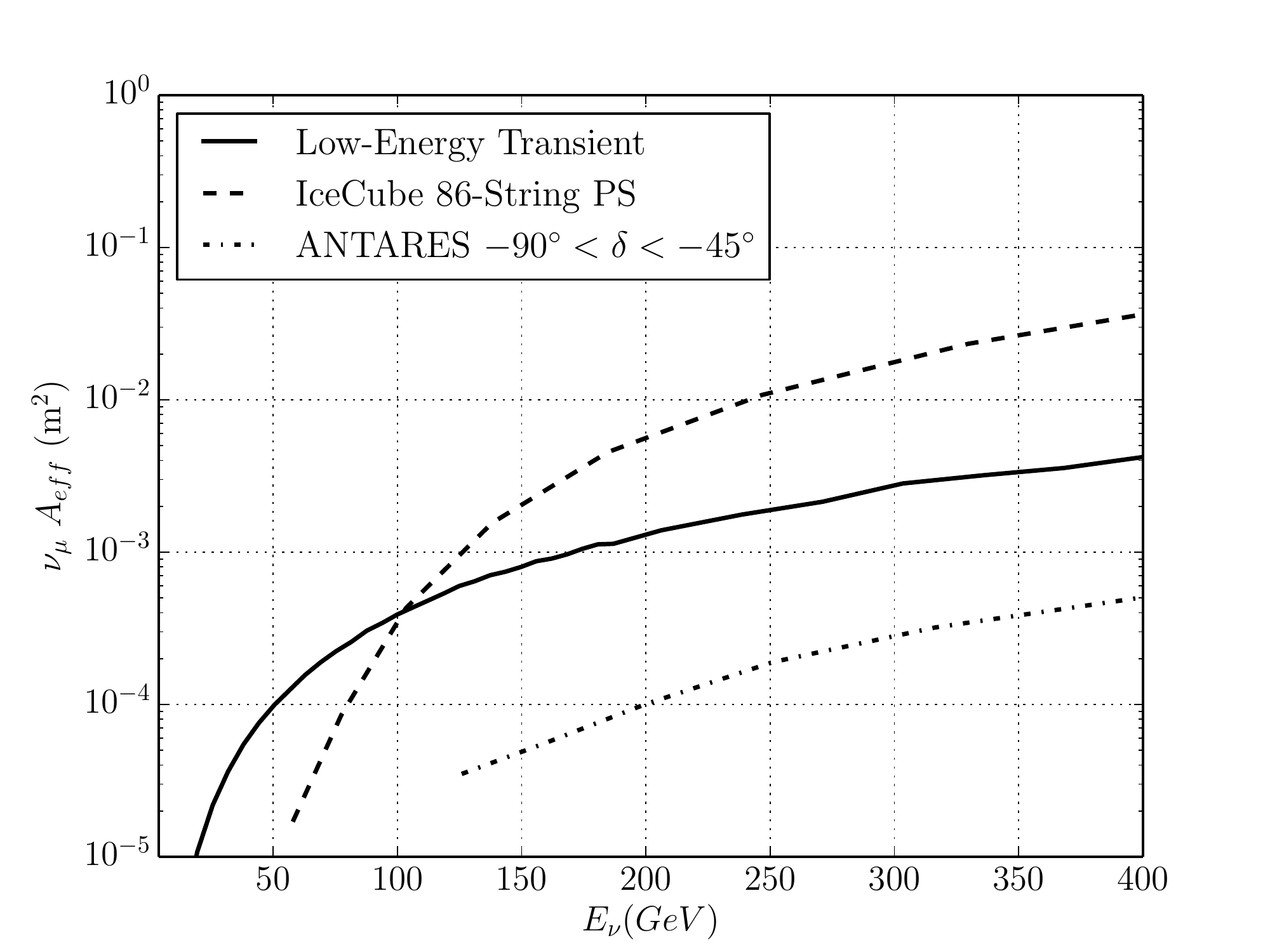}{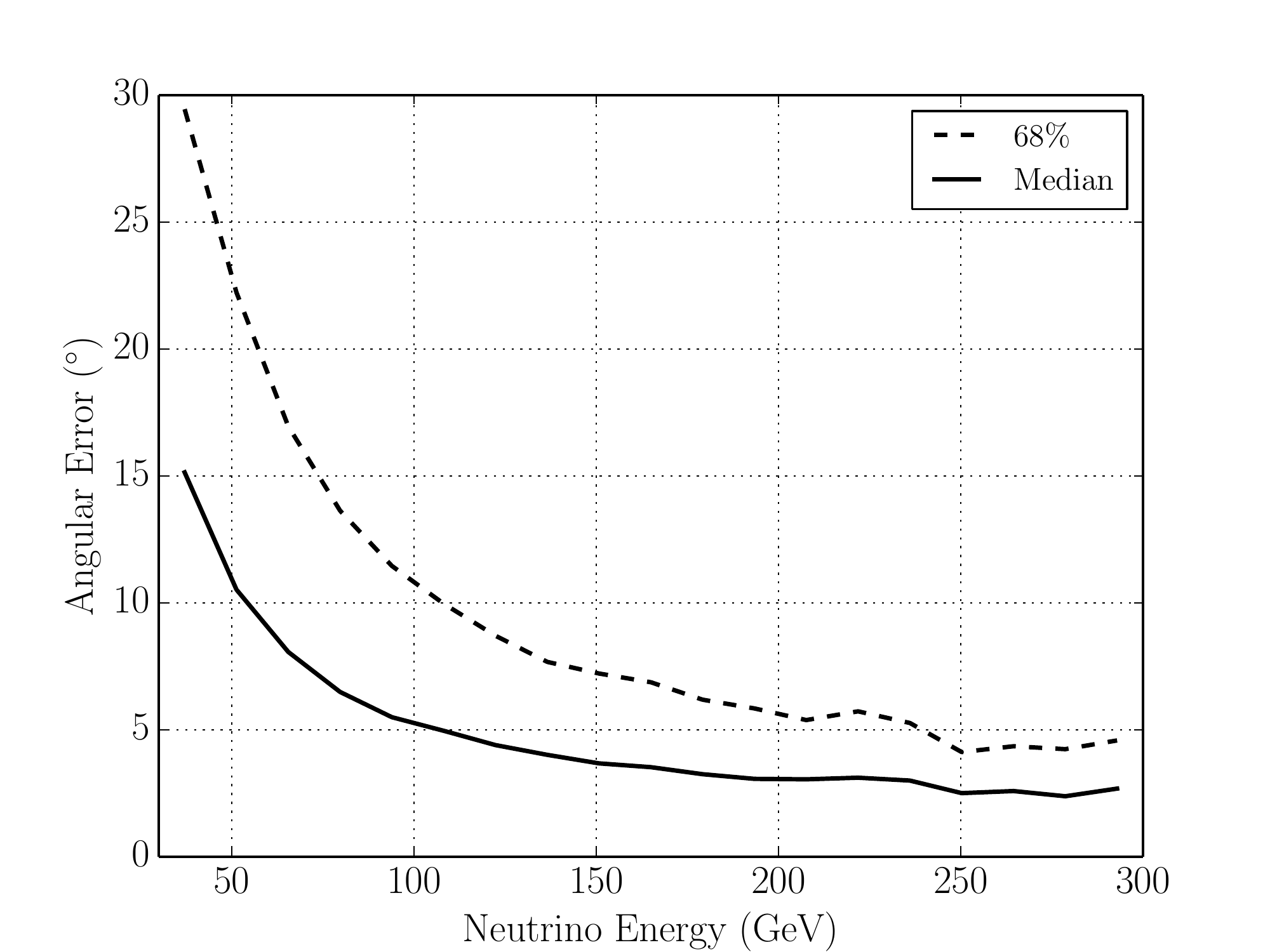}
\caption[Analysis Effective Area and Angular Resolution]{(left) The muon neutrino effective area as a function of neutrino energy for the presented search. The effective areas for both the 4 year IceCube point source search \citep{2014ApJ...796..109A} and the 4 year ANTARES point source search \citep{2012ApJ...760...53A} are plotted as well for comparison. (right) Muon neutrino angular resolution as a function of energy after event selection.}
\label{fig:EffAreaAndAngularResolution}
\end{figure}

The neutrino effective area for this event selection is shown in Figure \ref{fig:EffAreaAndAngularResolution}. While standard IceCube analyses clearly have superior sensitivity at higher energies, this event selection shows increased acceptance for events below about 100 GeV in neutrino energy. Figure \ref{fig:EffAreaAndAngularResolution} also shows the angular resolution for events at the analysis level as a function of energy. Lower neutrino energies result in muon tracks that are both shorter and dimmer, leading to difficulty in resolving the direction of the neutrino primary. The kinematic angle between the neutrino primary and muon secondary also contributes to the angular error. The median kinematic muon-neutrino angle after event selection ranges from $\sim$3$^{\circ}$ at 50 GeV to $\sim$1$^{\circ}$ at 300 GeV. As Figure \ref{fig:EffAreaAndAngularResolution} shows, the efficacy of the reconstruction method used in this analysis begins to deteriorate rapidly below 30 GeV due to too few DOM hits. Although the pointing ability of these low-energy neutrino events is limited, they are still able to contribute to the search through temporal correlation with other events in the sample.

\section{Analysis Method}
The search methods employed in the analysis of this data are nearly identical to those used in previous time-dependent IceCube analyses (see \cite{2008APh....29..299B} and \cite{0004-637X-807-1-46}). The arrival times and directions of events within the dataset are input to a likelihood function which is then used to perform a likelihood ratio test to compare a signal plus background hypothesis for the data to the background only hypothesis.

Construction of this likelihood function begins with the assignment of individual event probabilities that reflect the likelihood of seeing an event $i$ with arrival time $t_i$, reconstructed direction $\mathbf{x}_i$, and angular uncertainty $\sigma_i$ given a hypothetical source located at $\mathbf{x}_s$ with strength $n_s$ having a Gaussian time profile with mean time $t_0$ and width $\sigma_w$.
\begin{equation}\label{eq:EventProb}
\mathcal{P}_i(\mathbf{x}_i,t_i,\sigma_i|\mathbf{x}_s,n_s,t_0,\sigma_w) = \frac{n_s}{n_{\mathrm{tot}}} \mathcal{S}_i + \left(1-\frac{n_s}{n_{\mathrm{tot}}}\right) \mathcal{B}_i
\end{equation}
The $\mathcal{S}_i$ and $\mathcal{B}_i$ terms listed in Eq. \ref{eq:EventProb} are the signal and background probability density functions (p.d.f.) respectively. The p.d.f.s used in this search differ slightly from those in previously reported searches in that they use no reconstructed energy information. The signal p.d.f. is given by
\begin{equation}
\mathcal{S}_i(|\mathbf{x}_i-\mathbf{x}_s|,t_i,t_o,\sigma_w,\sigma_i) = S_i(|\mathbf{x}_i-\mathbf{x}_s|,\sigma_i) \cdot T_i(t_i,t_o,\sigma_w),
\end{equation}
where 
\begin{equation}
S_i(|\mathbf{x}_i-\mathbf{x}_s|,\sigma_i) = \frac{\kappa}{4\pi \sinh \kappa} \exp \left(\kappa \cos |\mathbf{x}_i-\mathbf{x_s}|\right)
\end{equation}
and
\begin{equation}
T_i(t_i,t_o,\sigma_w) = \frac{1}{\sqrt{2\pi}\sigma_w} \exp \left(-\frac{(t_i-t_o)^2}{2 \sigma_w^2}\right)
\end{equation}
The spatial component of the signal p.d.f., $S_i$, is the Kent-Fisher distribution \citep{Fisher_Bingham}, and it represents a slight deviation in the signal p.d.f. definition with respect to previous searches (see \cite{2014ApJ...796..109A}). This function is analogous to a 2-dimensional Gaussian distribution, but it is normalized to the 2-sphere rather than an infinite plane. The concentration parameter $\kappa$ is determined by the event angular uncertainty and is defined as $\kappa = \sigma_{i}^{-2}$. The temporal component of the signal p.d.f., $T_i$, is simply a Gaussian with mean emission time of $t_o$ and a width of $\sigma_w$.

The background p.d.f., $\mathcal{B}_i$, is derived from the final level data set which is dominated by background. It has the following form
\begin{equation}
\mathcal{B}_i(\mathbf{x}_i,t_i) = P_{BkgDec}(\delta_i)\frac{P_{BkgAz}(\alpha_i)}{T},
\end{equation}
where $T$ is the total livetime of the search, $P_{BkgDec}(\delta_i)$ is a p.d.f. describing the event declination distribution, and $P_{BkgAz}(\alpha_i)$ is a p.d.f. describing the event distribution in detector azimuth. These p.d.f.s are generated directly from data, without reference to background simulations.

The likelihood function itself is simply the product sum of all individual event probabilities:
\begin{equation}\label{eq:LLH}
\mathcal{L}(\mathbf{x}_s,n_s,t_0,\sigma_w) = \prod \mathcal{P}_i(|\mathbf{x}_i-\mathbf{x}_s|,n_s,t_i,t_0,\sigma_w,\sigma_i)
\end{equation}
The ratio between the likelihood function values under the background only hypothesis ($n_s=0$) and the signal plus background hypothesis is maximized by varying the source parameters $n_s$, $\sigma_w$, and $t_0$. The test statistic $\hat{\lambda}$ is then defined as the maximum value of the likelihood ratio:
\begin{equation}\label{eq:TS}
\hat{\lambda} = -2\log \left[\frac{\sqrt{2\pi}\hat{\sigma}_w}{T}\frac{\mathcal{L}(n_s = 0)}{\mathcal{L}(\mathbf{x}_s,\hat{n}_s,\hat{t}_o,\hat{\sigma}_w)} \right]
\end{equation}
with $\mathcal{L}(n_s = 0)$ corresponding to the likelihood of the null hypothesis and $\mathcal{L}(\mathbf{x}_s,{n}_s,\hat{t}_o,\hat{\sigma}_w)$ the likelihood of the signal plus background hypothesis with the best-fit values of the source parameters. Because this is a search for sources of finite duration over a limited timescale, the number of potential short duration flares within the data set exceeds that of flares of longer duration, leading to an effective trials factor. This results in a bias towards flares of shorter duration. We counteract this effect by introducing a marginalization term $T/\sqrt{2\pi}\hat{\sigma_w}$ in the test statistic formulation which serves to penalize flares of shorter duration. This term also ensures that the test statistic will asymptotically follow a $\chi^2$ distribution with degrees of freedom corresponding to the number of fitted parameters for data consisting solely of background events. More details about this term and its justification can be found in \cite{2010APh....33..175B}.

The $\chi^2$ behavior of the test statistic enables the maximized value $\hat{\lambda}$ to be used to estimate the pre-trials p-value of the best-fit flare through the invocation of Wilks's theorem \citep{wilks1938}. Because this search attempts to maximize the signal hypothesis over the whole Northern sky many times, the actual significance of a given flare must be adjusted to account for the effective number of trials accrued during the sky scan. We use the procedure detailed in \cite{0004-637X-807-1-46} that involves scrambling the event arrival times in the final dataset, which also serves to scramble the event right ascension. The search is performed on the randomized background data set and the p-value of the most significant flare in the search is recorded. Many iterations are performed to build a distribution of p-values which can then be compared to the p-value of the result from the real data. The fraction of background trials that result in a p-value of equal or greater significance than the observed p-value dictates the probability that the observed result is simply the consequence of a random background fluctuation. This probability is referred to as the post-trials p-value and it represents the true significance of the search result with proper trials factor correction.

In order to preserve generality, the presented search makes no use of information outside of the data set to designate source regions or time periods of interest. Instead, each point in the sky over a declination band ranging from -5$^{\circ}$ to 90$^{\circ}$ is examined. This is accomplished by discretizing the sky into separate bins and letting the location of these bins serve as a grid over which to test a hypothetical flaring source. As this is an unbinned likelihood analysis, the data itself is not binned and events may contribute to the likelihood at any location being tested. Maximization of the likelihood is then performed to obtain a test statistic $\hat{\lambda}$ for each grid point. The first iteration of this scan uses a relatively coarse 2$^{\circ}$ by 2$^{\circ}$ binning. After this first scan, a followup scan with finer 0.5$^{\circ}$ by 0.5$^{\circ}$ binning is performed over the coarse bins featuring a pre-trials p-value more significant than a predefined threshold ($-\log_{10}($p-value$) > 1.75$). The result is a map of pre-trials p-values which shows the estimated significance of the best-fit flare hypothesis at each grid point in the scan. The best-fit flare from the point featuring the most significant maximized test statistic after both scans is returned as the hottest spot in the search.

\section{Results and Interpretations}
\begin{figure}[ht]
  \begin{center}
    \includegraphics[width=1.0\textwidth,keepaspectratio]{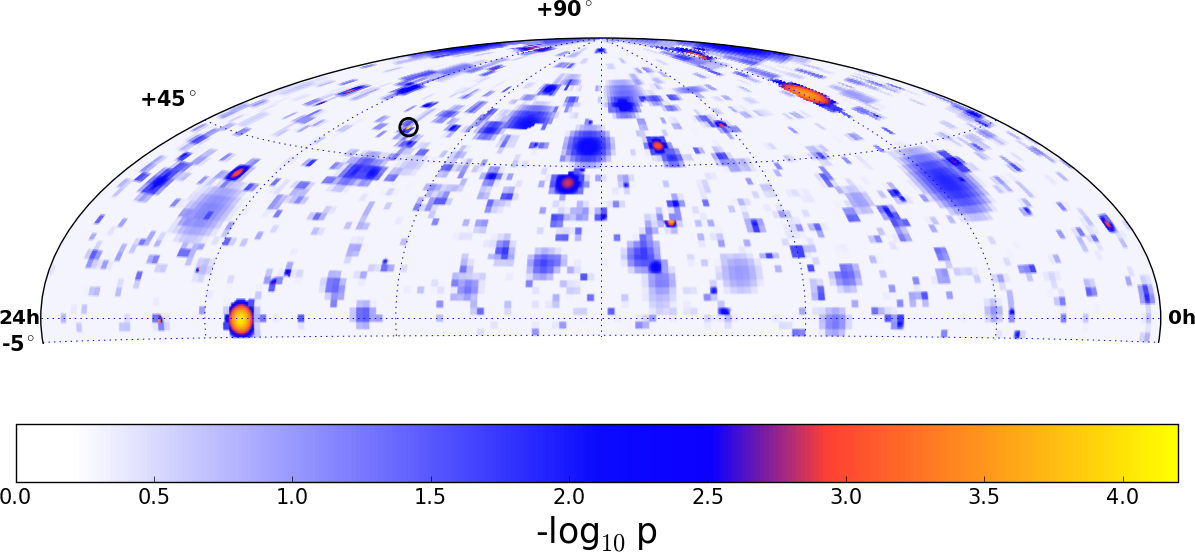}
  \end{center}
  \caption[Results Sky Map]{Sky map of pre-trials p-values for best fit flares per bin. The black circle identifies the location of the most significant flare found at RA = 268.75$^\circ$ and Declination = 54.25$^\circ$. The nominal resolution of the map is 2$^{\circ}$, however regions with more significant p-values receive a finer 0.5$^{\circ}$ resolution.}
  \label{fig:RealSkyMap}
\end{figure}

\begin{figure}[ht]
  \begin{center}
    \includegraphics[width=0.7\textwidth,keepaspectratio]{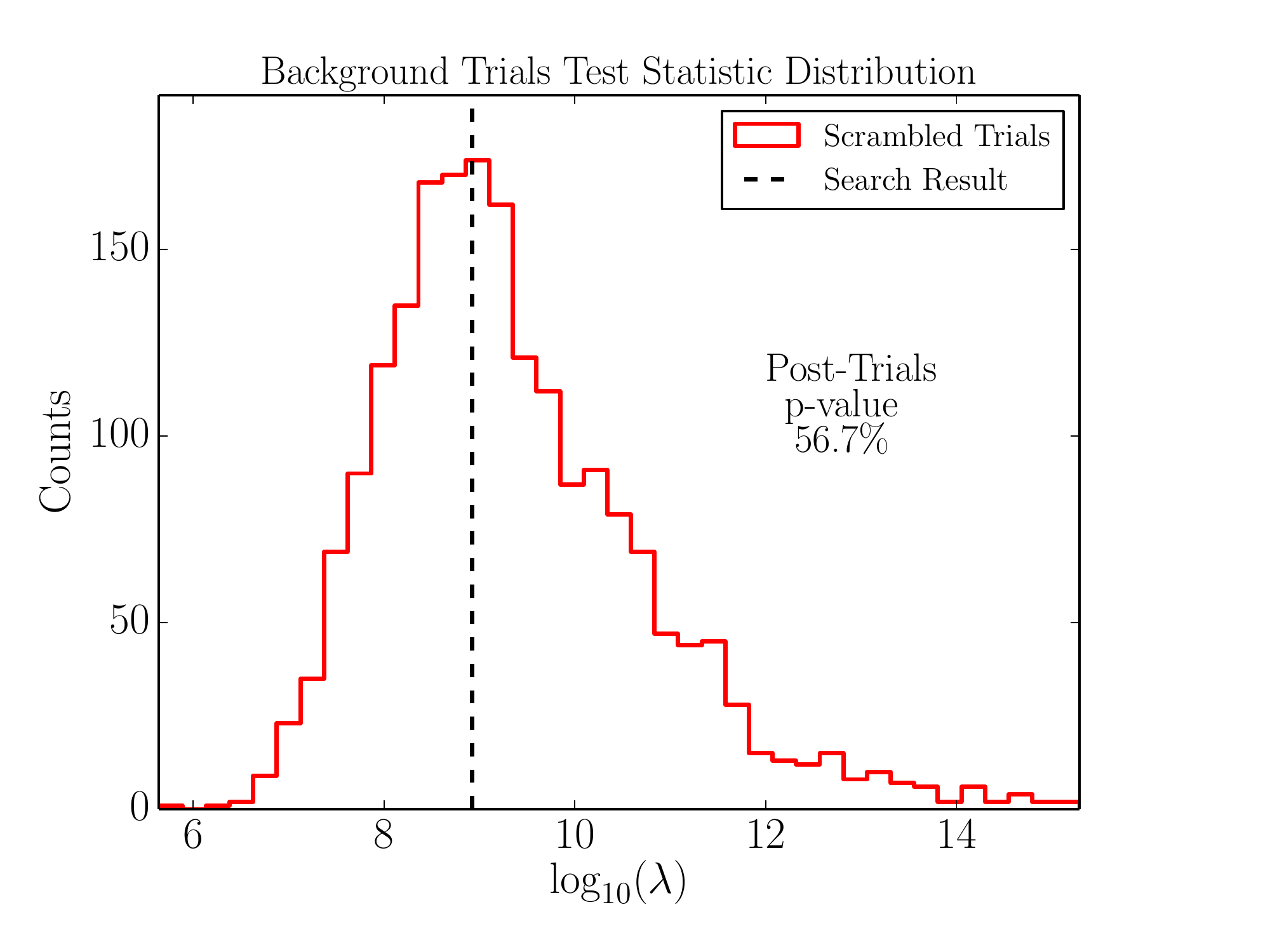}
  \end{center}
  \caption[Background Test Statistic Distribution]{Distribution of maximized test statistic $\hat{\lambda}$ for $2\times 10^4$ searches performed on randomized datasets. The dashed line indicates the value of $\hat{\lambda}$ for the most significant flare found in the data.}
  \label{fig:BkgTSDistribution}
\end{figure}

Applying the described analysis method to the unscrambled dataset yields the skymap of the pre-trials p-values shown in Figure \ref{fig:RealSkyMap}. The most significant flare is located at (RA, Dec.) = (268.75$^{\circ}$, 54.25$^{\circ}$) with a signal strength $n_s$ of 13.53 signal events and a width $\sigma_w$ of 5.89 days with the peak occurring on MJD 56107 (2012 June 29). The pre-trials p-value for this flare is estimated at 6.68$\times 10^{-5}$. The test statistic for this flare is compared to a background test statistic distribution constructed from $2 \times 10^4$ scrambled trials in Figure \ref{fig:BkgTSDistribution}. The background distribution gives a post-trials probability of seeing such a flare in a data set consisting of background only is 56.7$\%$, indicating that this flare is entirely consistent with the background hypothesis of the data.
In light of this null result, we can set an upper limit on the time-integrated neutrino flux of any possible unobserved neutrino flare that may have occurred during the search period.
\subsection{Generic Source Limit}
Due to the focus on low-energy events in this search, we choose to examine the limit with respect to a soft-spectrum $E^{-3}$ generic flaring neutrino source with a Gaussian emission profile. An upper limit is established through signal injections at a specified location through the following process. First, we select the p-value of the most significant flare found in the data to serve as a threshold for signal injection trials. Signal events are then injected with some Poisson mean value that is increased until the recovered p-values from the injections exceed the threshold p-value 90$\%$ of the time. This Poisson mean number of signal events is then taken as an event upper limit for the analysis method.

\begin{figure}[ht]
\plottwo{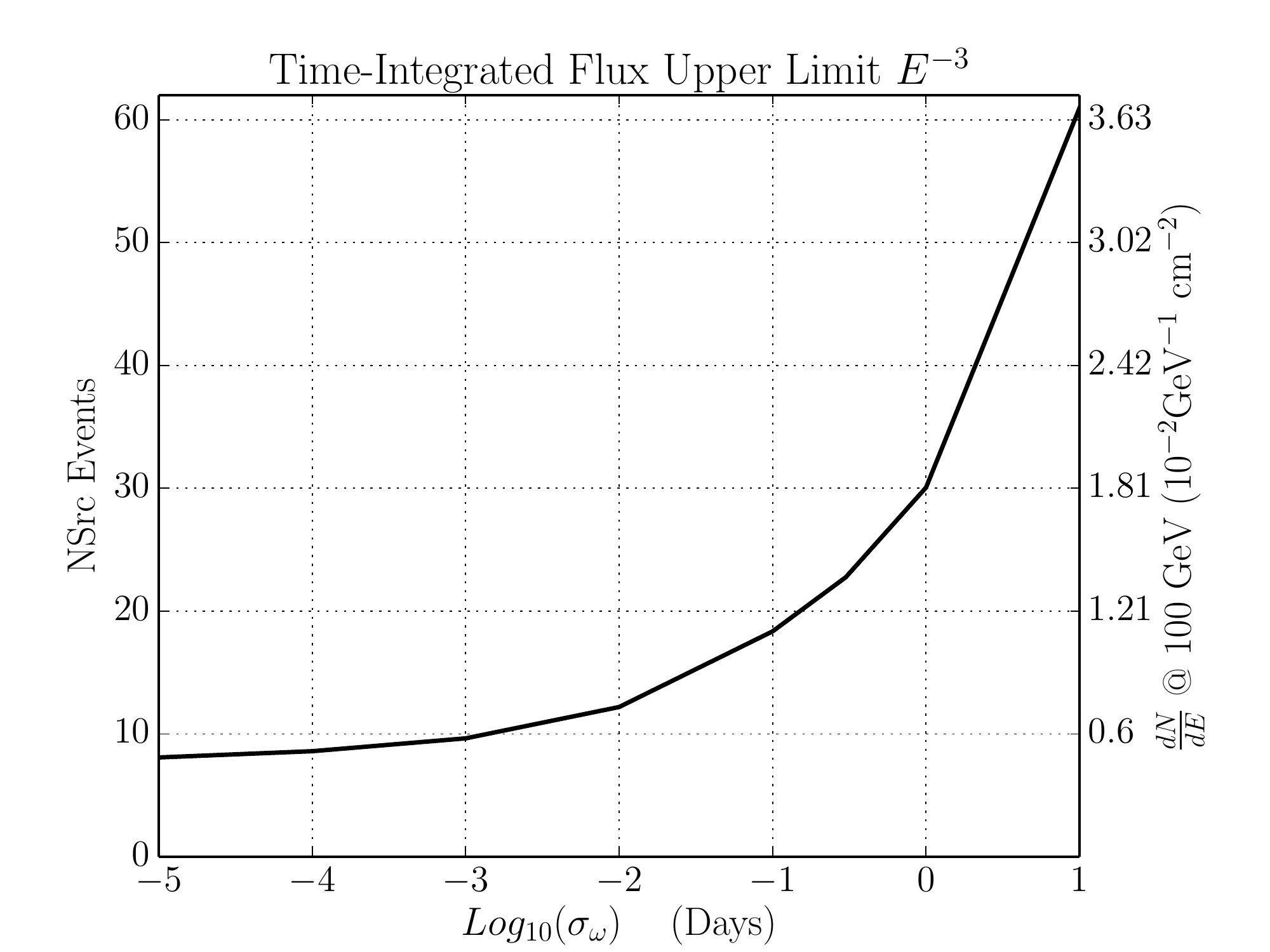}{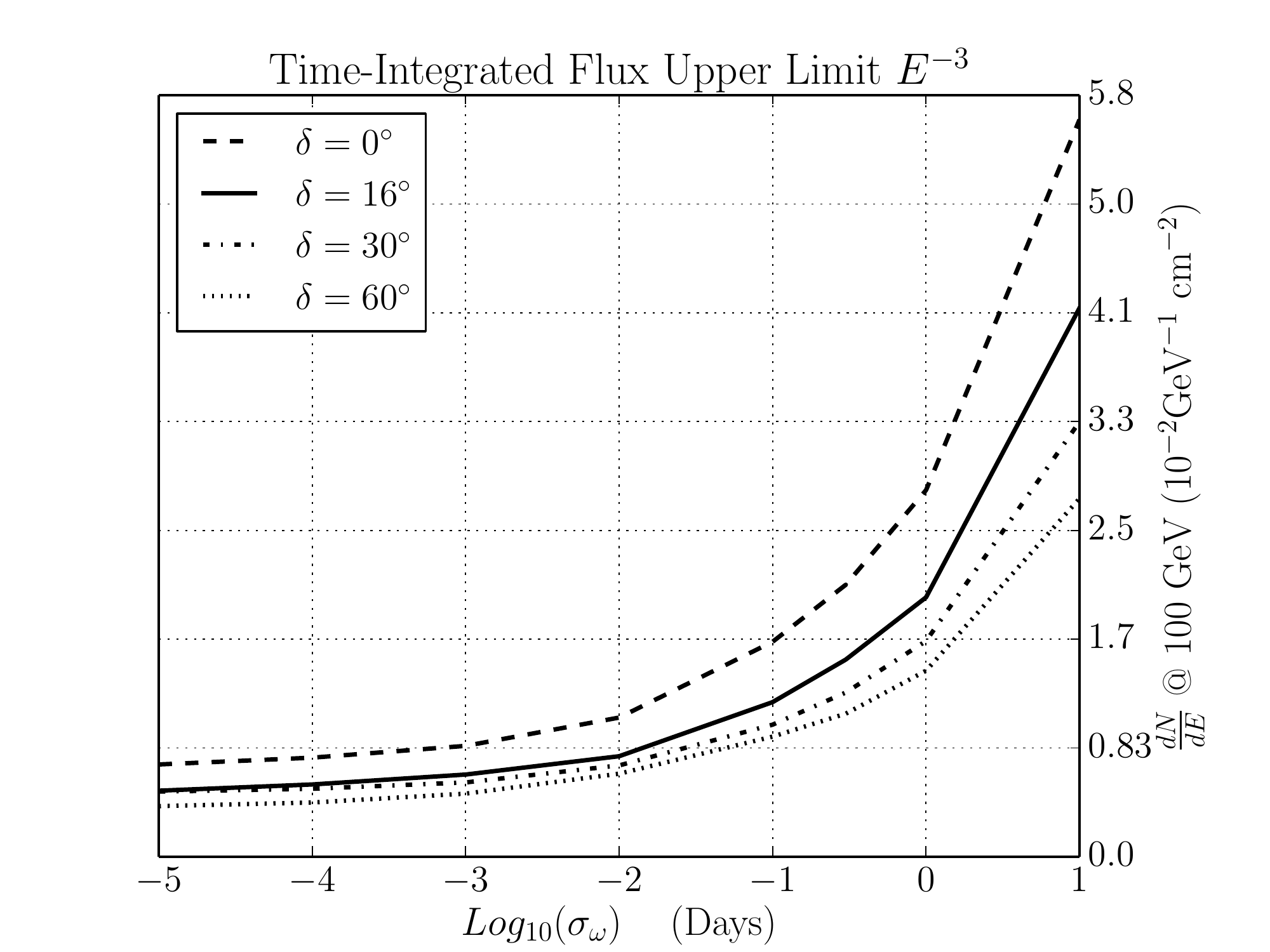}
\caption[Time-integrated Flux Limit for E$^{-3}$ Source]{(left) Upper limit (90$\%$ C.L.) for a generic E$^{-3}$ transient source as a function of flare width $\sigma_w$ averaged over the declination range of the analysis. The limit is given in mean number of events (left axis) as well as in time-integrated flux at a reference energy of 100 GeV (right axis). (right) Upper limit (90$\%$ C.L.) for a generic E$^{-3}$ transient source as a function of flare width $\sigma_w$ for different values of source declination.}
\label{fig:GenericE3Limit}
\end{figure}

The upper limit on a generic flaring source for several emission timescales and choices of declination is plotted in Figure \ref{fig:GenericE3Limit}. The number of events required rises at longer timescales as the rate of accidental background correlations becomes non-negligible. The limit in terms of time-integrated flux (GeV$^{-1} \cdot$ cm$^{-2}$) is also plotted. This limit is obtained by folding the source spectrum with the effective area of the event selection and normalizing the flux so that the number of events produced in the detector corresponds to the calculated Poisson mean event upper limit.

\subsection{Choked GRB Limits}
This null result can also be used to construct limits on specific neutrino emission models such as the RMW/AB model for choked GRB emission mentioned previously. Unlike the hard spectra sources (e.g, E$^{-2}$) that are the typical target in IceCube searches, the neutrino flux for choked GRBs is predicted to be much softer. The spectral shape can be modeled via a doubly broken power law with spectral breaks occurring as hadronic ($E_{\nu^{(1)}}$) and radiative ($E_{\nu^{(2)}}$) cooling mechanisms become efficient (see Eq. \ref{eq:chkgrb_spec}). Using the canonical RMW/AB model parameters, the break energies for pions (kaons) occur at 30 GeV (200 GeV) and 100 GeV (20 TeV). Therefore the neutrino spectrum is predicted to be very soft at $\gtrsim 1$ TeV energies.
\begin{equation}\label{eq:chkgrb_spec}
\frac{d\Phi_\nu}{dE}=F_\nu\left\{\begin{array}{cc}
E^{-2} & E < E_{\nu}^{(1)} \\ 
E_{\nu}^{(1)}E^{-3} & E_{\nu}^{(1)}< E < E_{\nu}^{(2)} \\ 
E_{\nu}^{(1)}E_{\nu}^{(2)}E^{-4} & E_{\nu}^{(2)}< E < E_{max}
\end{array}\right.
\end{equation}
\begin{equation}\label{eq:nuflu}
F_{\nu} = \frac{<n>_{\pi(K)}B_{\pi(K)}}{8} \cdot \frac{E_j \Gamma_b^2}{2 \pi D^2 \textrm{ln}(E_{p,max}'/ E_{p,min}')}
\end{equation}
The fluence $F_\nu$ at the Earth is given by Eq. \ref{eq:nuflu} and depends upon the pion (kaon) multiplicity $<n>$, the neutrino production branching ratio for pions (kaons) $B_{\pi(K)}$, the minimum and maximum proton energies ($E_{p,min}', E_{p,max}'$), the kinetic energy of the jet $E_j$, the bulk Lorentz factor $\Gamma_b$, and lastly the distance to the source $D$. Equations \ref{eq:chkgrb_spec} and \ref{eq:nuflu} reveal that the normalization of the neutrino flux at the Earth is highly dependent on the kinetic energy of the jet $E_j$ and the bulk lorentz factor $\Gamma_b$. These two parameters also determine the shape of the spectrum as the hadronic ($E_{\nu^{(1)}}\propto E_j^{-1} \Gamma_b^5$) and radiative ($E_{\nu^{(2)}}\propto \Gamma_b$) break energies depend upon these jet properties as well. We therefore choose to examine the predicted neutrino fluence in $E_j$-$\Gamma_b$ phase space. 

To determine which values of these parameters produce a fluence detectable through our search method, an event upper limit is first determined via the injection of signal events following a spectrum set by the value of $E_j$ and $\Gamma_b$ (the same process used to calculate event sensitivity for the generic $E^{-3}$ scenario). The emission timescale for these injections is given by a Gaussian with a width of 100\,s. The calculated event upper limit is then combined with the effective area of the event selection to determine the neutrino fluence necessary for detection. For a given choice of $E_j$ and $\Gamma_b$ this sets a limit on the distance at which the source would still be visible to the search, and we define this distance $D_{vis}$ as the visibility distance.

When combined with the area of sky examined by the search $\Omega_{A}$, this visibility distance, in turn, defines a parameter dependent volume $V_A$ ($=\frac{1}{3}\Omega_{A}D_{vis}^{3}$) over which the search method monitors. This monitored volume corresponds to the region in which a choked GRB event should be visible to the presented search method with 90$\%$ confidence (assuming jet alignment). If the observation period of the search is considered, this monitored volume can be converted into a limit on the volumetric rate of choked GRB events as a function of $E_j$ and $\Gamma_b$. This, however, requires the assumption that the jets of any choked GRB event in this volume are aligned with the Earth. To obtain a limit more representative of the actual distribution of choked GRB orientations, one can include a geometrical correction factor that takes into account the opening angle of the jets, $\theta_j$, which is often approximated as $\theta_j \sim \frac{1}{\Gamma_b}$ \citep{2013ApJ...777..162M}. Because the physics that determines this opening angle is not entirely known, we choose not to include any correction for jet opening angle. The rate limit is then given by
\begin{equation}\label{eq:ratelimit}
R = \left(\frac{U.L.(0|\mu)}{\tau \cdot V_A}\right),
\end{equation}
where $\tau$ is the livetime of the search, $V_A$ is the monitored volume previously defined, and $U.L.(0|\mu)$ is the null observation upper limit on the number of choked GRBs that occurred in our monitored volume with background expectation of $\mu$.

We define this background expectation $\mu$ as the expected number of `false positive' flares that occur due to coincident background events during a given search. To calculate the value of $\mu$, we first perform many iterations of the analysis with $n_s$ signal events injected at a specific declination where $n_s$ is the calculated event sensitivity at that declination. The test statistic for these injection trials form a distribution from which we can take the median value, $\lambda_{inj}^{med}$. Once $\lambda_{inj}^{med}$ has been determined, the analysis is run again scanning over the same declination band using a time-scrambled dataset with no injections. Several iterations of this procedure builds a background test statistic distribution. The number of entries in the background distribution whose test statistic value exceeds the threshold $\lambda_{inj}^{med}$ are then recorded. This number is then divided by the number of background-only analysis iterations performed to yield an expected false positive rate per search. This procedure revealed the false positive rate to be very small ($\leq 10^{-3}$). We therefore take $\mu\approx0$ leading to a Neyman upper limit of 2.3 from the null observation.
\begin{figure}[ht]
  \begin{center}
    \includegraphics[width=0.8\textwidth,keepaspectratio]{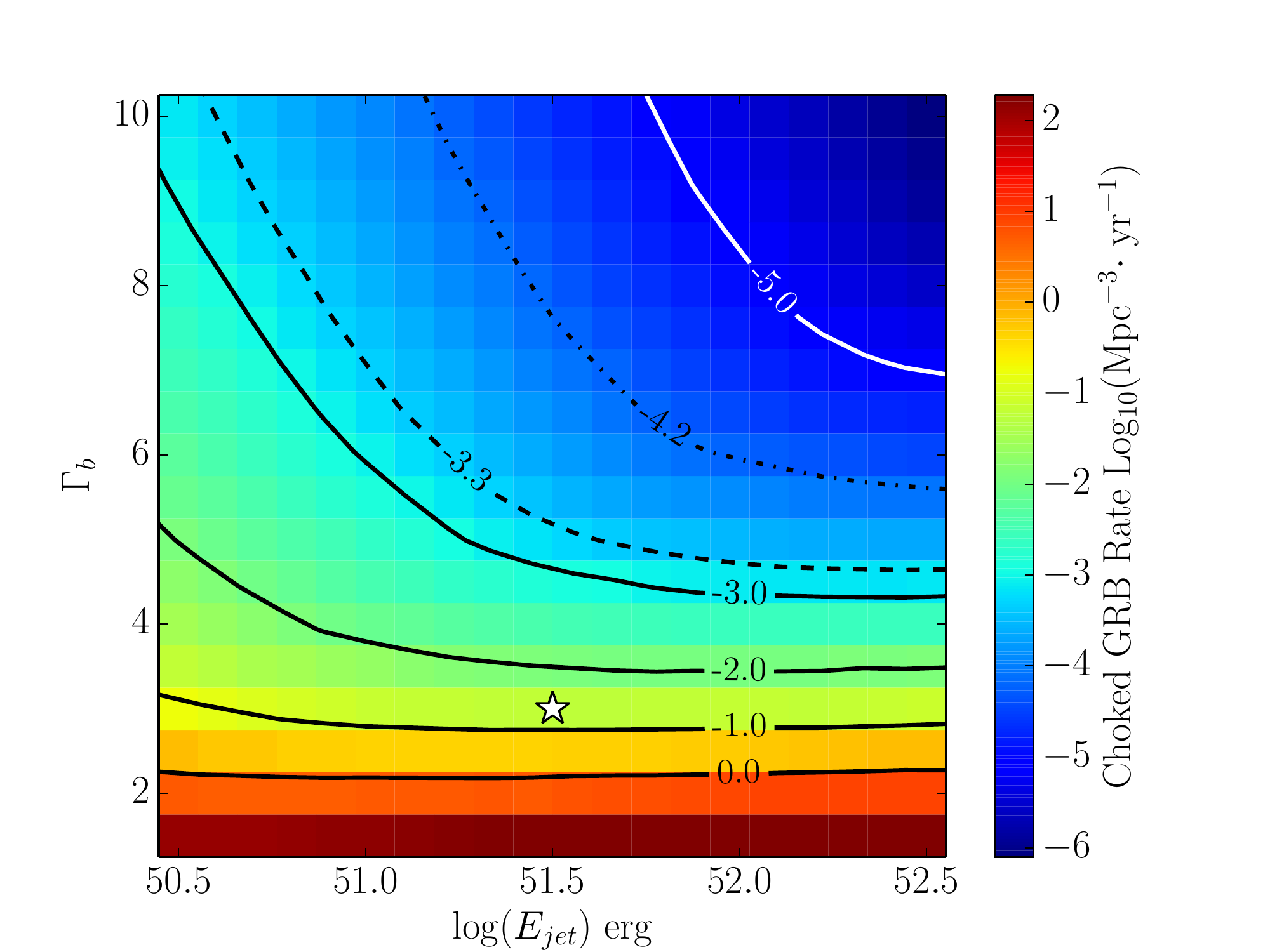}
  \end{center}
  \caption[Choked GRB Volumetric Rate Limit]{Plot of the volumetric rate limit on choked GRBs in the nearby universe. The bin for canonical values of the RMW/AB emission model is marked by the star. The dashed line contour gives the rate of core-collapse supernovae within 10 Mpc as measured by \cite{2011PhRvD..83l3008K}. The dot-dashed line is the volumetric rate extracted from a large survey of SNe in the local universe \citep{2011MNRAS.412.1419L}.}
  \label{fig:VolumetricRateSensitivity}
\end{figure}

The volumetric rate limit for a range of values of $E_j$ and $\Gamma_b$ is plotted in Figure \ref{fig:VolumetricRateSensitivity}. Two separate measurements of the nearby CC SNe are plotted as well to provide context to the calculated rate limits. Choked GRB events harboring particularly energetic jet parameters should be visible to the search method. However, if one compares the limits for the canonical RMW/AB model parameter values ($\Gamma_b = 3$, $E_j = 10^{51.5}$ erg) to the CC SNe rates, it is clear that the current search method is not very sensitive to large regions of the model parameter space. However, the sensitivity of this search can be improved through refinement of the event selection and analysis methods. Potential changes include greater signal retention through more efficient use of multi-variate machine learning cuts in the event selection process, the use of reconstruction methods optimized for sub-TeV muon tracks, and more accurate modeling of event angular error distribution. 

\section{Conclusions}
The described search examined a newly developed data set consisting of 30-300 GeV muon neutrinos. No evidence for transient astrophysical neutrino sources was found in the data, leading to the construction of upper limits on the neutrino fluence of potential sources within the observation period. In particular, we examine the derived limit in the context of neutrino emission from choked GRBs. Although this search in its current configuration is only sensitive to particularly energetic or nearby choked GRBs, the sensitivity of this method will improve as the event selection and search techniques are further optimized for muon neutrino events at sub-TeV energies. Continued development of this event selection will complement the current mature IceCube analyses at higher energies, leading to an overall enhancement of the detector's sensitivity to transient sources.

\acknowledgments
We acknowledge the support from the following agencies:
U.S. National Science Foundation-Office of Polar Programs,
U.S. National Science Foundation-Physics Division,
University of Wisconsin Alumni Research Foundation,
the Grid Laboratory Of Wisconsin (GLOW) grid infrastructure at the University of Wisconsin - Madison, the Open Science Grid (OSG) grid infrastructure;
U.S. Department of Energy, and National Energy Research Scientific Computing Center,
the Louisiana Optical Network Initiative (LONI) grid computing resources;
Natural Sciences and Engineering Research Council of Canada,
WestGrid and Compute/Calcul Canada;
Swedish Research Council,
Swedish Polar Research Secretariat,
Swedish National Infrastructure for Computing (SNIC),
and Knut and Alice Wallenberg Foundation, Sweden;
German Ministry for Education and Research (BMBF),
Deutsche Forschungsgemeinschaft (DFG),
Helmholtz Alliance for Astroparticle Physics (HAP),
Research Department of Plasmas with Complex Interactions (Bochum), Germany;
Fund for Scientific Research (FNRS-FWO),
FWO Odysseus programme,
Flanders Institute to encourage scientific and technological research in industry (IWT),
Belgian Federal Science Policy Office (Belspo);
University of Oxford, United Kingdom;
Marsden Fund, New Zealand;
Australian Research Council;
Japan Society for Promotion of Science (JSPS);
the Swiss National Science Foundation (SNSF), Switzerland;
National Research Foundation of Korea (NRF);
Danish National Research Foundation, Denmark (DNRF)

\bibliography{ms}{}

\end{document}